\documentclass{article}
\usepackage[utf8]{inputenc}
\usepackage{bm}
\usepackage[svgnames]{xcolor}
\usepackage[colorlinks=true,linkcolor=DarkBlue,citecolor=DarkBlue,anchorcolor=DarkBlue,urlcolor=DarkBlue]{hyperref}
\usepackage{amsmath}

\title{Binned Likelihood with Monte Carlo Statistical Uncertainty in Bayesian Inference}
\author{Shilin Liu, Clark McGrew }
\date{January 2023}

\begin{document}

\maketitle

\begin{abstract}

Data analysis in HEP experiments often uses binned likelihood from data and finite Monte Carlo sample. Statistical uncertainty of Monte Carlo sample has been introduced in Frequentist Inference in some literatures, but they are not suitable for Bayesian Inference. This technical note introduces the binned likelihood with Monte Carlo statistical uncertainty in Bayesian Inference and includes the derivation of it. It turns out that the results are similar to the results in \cite{arguelles2019binned}. But this tech-note gives an alternate and more intuitive derivation of the content.

\end{abstract}

\section{Extended Binned Likelihood}
Bayes's theorem is often used to obtain the data evidenced theory parameters distribution, as shown in equation (\ref{bayes' theorem}) :
\begin{equation}\label{bayes' theorem}
    P(\bm{\theta}|\bm{x})\;=\;\frac{P(\bm{x}|\bm{\theta})P(\bm{\theta})}{P(\bm{x})}
\end{equation}
In the equation, $\bm{\theta}$ represents the model parameters and $\bm{x}$ represents the data obtained from the experiment. 
\newline

$P(\bm{\theta})$ is the prior distribution of $\bm{\theta}$, which is already known from our prior knowledge. $P(\bm{x})$ can be obtained using the law of total probability and is a constant, which can be ignored in most sampling processes.
\newline

$P(\bm{x}|\bm{\theta})$ is the likelihood and is what will be discussed in this document. Denoting the number of events in the ith bin from data as $N^i_{data}$, the number of events in the ith bin from Monte Carlo sample with model parameters $\bm{\theta}$ as $N^i_{mc}$, and the Monte Carlo to data POT scale factor $s$. Then $P(\bm{x}|\bm{\theta})$ can be calculated using binned data likelihood outlined by G. Cowan\cite{1}, where the number of data events in each bin is taken as to follow a Poisson distribution with expectation to be $N^i_{mc}/s$ (model prediction). Therefore, the likelihood $P(\bm{x}|\bm{\theta})$ can be expressed as equation (\ref{joint poisson likelihood})

\begin{equation}\label{joint poisson likelihood}
  P(\bm{x}|\bm{\theta}) =\prod_{i=1}^{nbins}\frac{{(N^{i}_{mc}/s)}^{N^i_{data}}}{N^i_{data}!}e^{-N^{i}_{mc}/s}
\end{equation}

Taking the natural log of equation (\ref{joint poisson likelihood}) and Stirling's approximation ($\ln(n!) = n\ln n - n + O(\ln n)$) gives following likelihood (the superscript i is omitted for simplicity), which has been widely used in the T2K collaboration.
\begin{equation}\label{log poisson likelihood}
  -\ln P(\bm{x}|\bm{\theta}) =\sum_{i=1}^{nbins} \left[ N_{mc}/s - N_{data} + N_{data} \ln \frac{N_{data}}{N_{mc}/s} \right]
\end{equation}

\section{MC statistical uncertainty treatment in Bayesian Inference}
In this section, all the discussion is based on a certain bin (it can be any bin).
\newline

Notice that in equation \ref{joint poisson likelihood}, an assumption was made that $N_{mc}$ is the truth expectation of number of data events. However, $N_{mc}$ is a Monte Carlo sample, which is a statistical fluctuation from the truth number of Monte Carlo events (could be obtained with infinite statistics, denoted by $N_{\bm{\theta}}$). And data is also a statistical fluctuation from the truth number of data events (denoted by $N_{true}$).
\newline

Therefore, the likelihood should be calculated based on data ($N_{data}$) and data truth ($N_{true}$), Monte Carlo ($N_{mc}$) and Monte Carlo truth ($N_{\bm{\theta}}$). From the law of total probability, one can obtain equation (\ref{data mc truth}).
\begin{equation}\label{data mc truth}
   P(x|\bm{\theta}) =\int P(x|N_{true}) \left[ \int P(N_{true}|N_{\bm{\theta}})P(N_{\bm{\theta}}|\bm{\theta})dN_{\bm{\theta}} \right] dN_{true}
\end{equation}

Since both $N_{true}$ and $N_{\bm{\theta}}$ represent the truth expectation of number of events (for data and Monte Carlo), it is natural that they should equal (with the POT correction). Therefore $P(N_{true}|N_{\bm{\theta}})$ can be expressed with a delta function (equation (\ref{delta})).
\begin{equation}\label{delta}
    P(N_{true}|N_{\bm{\theta}}) = \delta (N_{true} - N_{\bm{\theta}} / s)
\end{equation}

The likelihood $P(x|N_{true})$ can be safely expressed following equation (\ref{joint poisson likelihood}), since $N_{true}$ is the truth expectation number of events in the bin (in data POT).
\begin{equation}\label{data truth}
   P(x|N_{true}) = \frac{N_{true}^{N_{data}}}{N_{data}!} e^{-N_{true}}
\end{equation}

The Monte Carlo simulation is often too complex and it is impossible to derive $N_{\bm{\theta}}$ from $\bm{\theta}$. Since $N_{mc}$ is a realization of parameter $\bm{\theta}$, it can be used together with Bayesian Inference to derive the term $P(N_{\bm{\theta}}|\bm{\theta})$

\begin{equation}\label{truth mc}
   P(N_{\bm{\theta}}|\bm{\theta}) = P(N_{\bm{\theta}}| N_{mc}, \bm{\theta}) = \frac{P(N_{mc}|N_{\bm{\theta}}) P(N_{\bm{\theta}})}{P(N_{mc})}
\end{equation}

In equation (\ref{truth mc}), since the Monte Carlo events are sampled for a finite period of time, its truth expectation ($N_{\bm{\theta}}$) can only contain a finite number of events. Therefore P($N_{\bm{\theta}}$) is taken as flat from zero to a large number, ub, compared to the possible Monte Carlo number of events, and does not introduce prior bias on the truth expectation of the number of Monte Carlo events. $P(N_{mc}|N_{\bm{\theta}})$ can be obtained similar as equation (\ref{data truth}).
\begin{equation}\label{mc truth}
   P(N_{mc}|N_{\bm{\theta}}) = \frac{(N_{\bm{\theta}})^{N_{mc}}}{N_{mc}!} e^{-N_{\bm{\theta}}}
\end{equation}

$P(\bm{\theta})$ can be obtained using the law of total probability and is a constant, as shown in equation (\ref{p(n_mc)}), in which \textit{ub} represents the integration upper bound.
\begin{equation}\label{p(n_mc)}
\begin{aligned}
    P(N_{mc}) 
    & = \int_{0}^{ub} P(N_{mc}|N_{\bm{\theta}}) P(N_{\bm{\theta}})dN_{\bm{\theta}}
    & = \int_{0}^{ub} \frac{(N_{\bm{\theta}})^{N_{mc}}}{N_{mc}!} e^{-N_{\bm{\theta}}} \frac{1}{ub} dN_{\bm{\theta}} = \frac{1}{ub}
\end{aligned}
\end{equation}

Plugging equations (\ref{delta}), (\ref{data truth}), (\ref{truth mc}), (\ref{mc truth}), and (\ref{p(n_mc)}) into equation (\ref{data mc truth}) gives equation (\ref{data parameter}). And the integration is approximated to a gamma function (since upper bound is large).
\begin{equation}\label{data parameter}
\begin{aligned}
   P(x|\bm{\theta}) 
    & =\int P(x|N_{true}) \left[ \int P(N_{true}|N_{\bm{\theta}})P(N_{\bm{\theta}}|\bm{\theta})dN_{\bm{\theta}} \right] dN_{true}\\
    & = \int \frac{N_{true}^{N_{data}}}{N_{data}!} e^{-N_{true}} \\
    & \; \left[ 
        \int_{0}^{ub} \delta (N_{true} - N_{\bm{\theta}} / s) 
        \frac{(N_{\bm{\theta}})^{N_{mc}}}{N_{mc}!} e^{-N_{\bm{\theta}}} \frac{1}{ub\;P(N_{mc})}dN_{\bm{\theta}}
        \right] dN_{true} \\
    & = \int_{0}^{ub} 
            \frac{(N_{\bm{\theta}}/s)^{N_{data}}}{N_{data}!} e^{-N_{\bm{\theta}}/s} 
            \frac{(N_{\bm{\theta}})^{N_{mc}}}{N_{mc}!} e^{-N_{\bm{\theta}}} 
        dN_{\bm{\theta}} \\
    & = \frac{1}{N_{data}!N_{mc}!s^{N_{data}}} 
        \int_{0}^{ub} 
            N_{\bm{\theta}}^{N_{data} + N_{mc}} 
            e^{-N_{\bm{\theta}}(1 + 1/s)}
        dN_{\bm{\theta}} \\
    & = \frac{1}{N_{data}!N_{mc}!s^{N_{data}}} 
        \frac{1}{(1+1/s)^{N_{data} + N_{mc} + 1}}\\
    & \; \int_{0}^{ub\; (1+1/s)}
            u^{N_{data} + N_{mc}}
            e^{-u}
          du\\
    & = \frac{1}{N_{data}!N_{mc}!s^{N_{data}}} 
        \frac{1}{(1+1/s)^{N_{data} + N_{mc} + 1}}
        \Gamma(N_{data} + N_{mc} + 1)\\
    & = \frac{(N_{data} + N_{mc})!}{N_{data}!N_{mc}!s^{N_{data}}} 
        \frac{1}{(1+1/s)^{N_{data} + N_{mc} + 1}}\\
    & = \frac{s}{(1+s)}
        \frac{(N_{data} + N_{mc})!}{N_{data}!N_{mc}!}
        (\frac{1}{1+s})^{N_{data}}
        (\frac{s}{1 + s})^{N_{mc}}
\end{aligned}
\end{equation}

Interestingly, the simplified likelihood is nothing but a binominal distribution, with total number of events $N_{data} + N_{mc}$, with two outcomes: (1) data, with a probability of 1/(1+s), and the number of times for this outcome is $N_{data}$. (2) mc, with a probability of s/(1+s), and the number of times for this outcome is $N_{mc}$.
\newline
Taking the natural log of equation (\ref{data parameter}) and Stirling's approximation ($\ln(n!) = n\ln n - n + O(\ln n)$) gives following likelihood, in which the constant terms are dropped.
\begin{equation}\label{log likelihood with MC statistical uncertainty}
\begin{aligned}
    - \ln P(x|\bm{\theta}) 
    & = -(N_{data} + N_{mc}) \ln (N_{data} + N_{mc}) + N_{data} + N_{mc}\\
    & \;\; + N_{data} \ln N_{data} - N_{data} \\
    & \;\; + N_{mc} \ln N_{mc} - N_{mc} \\
    & \;\; + N_{data} \ln{s} \\
    & \;\; + (N_{data} + N_{mc} + 1) \ln (1 + 1/s)\\
    & = N_{data} \ln \frac{s\;N_{data}}{N_{data} + N_{mc}} - N_{mc} \ln \frac{N_{data} + N_{mc}}{N_{mc}}\\
    & \;\; + (N_{data} + N_{mc} + 1) \ln (1 + 1/s)\\
    & = (N_{mc} + N_{data} + 1) \ln (1 + 1/s) 
        - N_{mc} \ln (1 + \frac{N_{data}}{N_{mc}}) \\
    & \;\; + N_{data} \ln \frac{N_{data}}{N_{mc}/s + N_{data}/s}
\end{aligned}
\end{equation}

In the case of a large Monte Carlo sample (s $\sim $ order of 10). Equation (\ref{log likelihood with MC statistical uncertainty}) can be approximated as
\begin{equation}\label{log likelihood with MC statistical uncertainty approximate }
\begin{aligned}
    - \ln P(x|\bm{\theta}) = 
    & \frac{1}{s} + (N_{mc}/s + N_{data}/s) - N_{data}
     + N_{data} \ln \frac{N_{data}}{N_{mc}/s + N_{data}/s} \\
\end{aligned}
\end{equation}

Taking one bin from equation (\ref{log poisson likelihood}) gives equation (\ref{log poisson likelihood one bin}). Notice that the major difference between equation (\ref{log likelihood with MC statistical uncertainty approximate }) and equation (\ref{log poisson likelihood one bin}) is the additional term $N_{data}/s$ to $N_{mc}/s$, which is small when s is large (Monte Carlo sample size is large).

\begin{equation}\label{log poisson likelihood one bin}
 -\ln P(x|\bm{\theta}) 
= N_{mc}/s - N_{data} + N_{data} \ln \frac{N_{data}}{N_{mc}/s} 
\end{equation}

\section{Comparison with Approximated Barlow-Beeston Likelihood}

In \cite{conway2011incorporating}, equation (8) introduces the approximated Barlow-Beeston likelihood. This section compares the difference between the Bayesian likelihood and the approximated Barlow-Beeston likelihood.
\newline

The approximated Barlow-Beeston likelihood is derived for Frequentist Inference where the maximum likelihood is needed \cite{barlow1993fitting}\cite{conway2011incorporating}. It makes some reasonable approximation (approximate equation (\ref{mc truth}) with a Gaussian distribution since Monte Carlo sample size is large). And in equation (\ref{data parameter}), instead of integrating over the truth number of Monte Carlo events ($N_{\bm{\theta}}$), it takes derivative with respect to $N_{\bm{\theta}}$ and finds the maximum likelihood. This is valid in Frequentist Inference since the maximum likelihood is wanted. 
\newline

However, in Bayesian Inference, this means the integration in equation (\ref{data parameter}) ($\int_{0}^{ub}\frac{(N_{\bm{\theta}}/s)^{N_{data}}}{N_{data}!} e^{-N_{\bm{\theta}}/s} \frac{(N_{\bm{\theta}})^{N_{mc}}}{N_{mc}!} e^{-N_{\bm{\theta}}}dN_{\bm{\theta}}$) is approximated by the peak of the approximated distribution ($\frac{(N_{\bm{\theta}}/s)^{N_{data}}}{N_{data}!} \frac{1}{\sqrt{2\pi} \sigma_{mc}} e^{-\frac{1}{2} (\frac{N_{\bm{\theta}} - N_{mc}}{\sigma_{mc}})^2}$). Generally in the sampling process, the difference (ration) between the likelihood with different model parameters $\bm{\theta}_1$ and $\bm{\theta}_2$ ($P(x|\bm{\theta}_1)$ and $P(x|\bm{\theta}_2)$) are needed to generate the distribution. If $N_{mc}$ remains a constant for different $\bm{\theta}$, with the condition that $\sigma_{mc}$ is also a constant, the approximated Barlow-Beeston likelihood can serve as a close approximation of the Bayesian likelihood.
\newline

But $N_{mc}$ always changes with different $\bm{\theta}$, since there is an accurate Bayesian likelihood available for Bayesian Inference, there's no need to bother with the approximations and use approximated Barlow-Beeston likelihood in Bayesian Inference.

\section{Conclusion}

The Bayesian log likelihood including Monte Carlo statistical uncertainty has been introduced and a closed form is given in equation (\ref{log likelihood with MC statistical uncertainty}). This likelihood is the valid likelihood for Bayesian Inference and it is comparable to the approximated Barlow-Beeston likelihood in Frequentist Inference.
\newline

It is interesting to notice that the Bayesian likelihood including Monte Carlo statistical uncertainty is a binominal distribution (equation (\ref{data parameter})).
\newline

It has also been shown that in the condition of large Monte Carlo sample size, the Bayesian likelihood including Monte Carlo statistical uncertainty approximates to the standard extended binned likelihood (equation (\ref{log likelihood with MC statistical uncertainty approximate }) and (\ref{log poisson likelihood one bin})).

\bibliographystyle{plain}
\bibliography{Reference.bib}

\begin{thebibliography}{1}

\bibitem{arguelles2019binned}
Carlos~A Arg{\"u}elles, Austin Schneider, and Tianlu Yuan.
\newblock A binned likelihood for stochastic models.
\newblock {\em Journal of High Energy Physics}, 2019(6):1--18, 2019.

\bibitem{barlow1993fitting}
Roger Barlow and Christine Beeston.
\newblock Fitting using finite monte carlo samples.
\newblock {\em Computer Physics Communications}, 77(2):219--228, 1993.

\bibitem{conway2011incorporating}
JS~Conway.
\newblock Incorporating nuisance parameters in likelihoods for multisource
  spectra.
\newblock {\em arXiv preprint arXiv:1103.0354}, 2011.

\bibitem{1}
G.~Cowan.
\newblock Statistical data analysis.

\end{thebibliography}
\end{document}